\documentclass[aps,amsmath,amssymb,superscriptaddress,showpacs,showkeys]{revtex4-2}

\usepackage[dvips]{graphicx}
\usepackage{times}
\usepackage{xcolor}
\usepackage{orcidlink}
\usepackage{hyperref}
\hypersetup{
  colorlinks=true,
  urlcolor=blue,
  linkcolor=red,
  citecolor=blue
}

\begin{document}


\title{AstroSat Observation of Recent Outburst in the Be/X-ray Binary LS V +44 
17/RX J0440.9+4431}

\author{Arshad Hussain}
\email{arshad007h@gmail.com}



\author{Umananda Dev Goswami\orcidlink{0000-0003-0012-7549}}
\email{umananda2@gmail.com}
\affiliation{Department of Physics, Dibrugarh University, Dibrugarh 786004, 
Assam, India}

\begin{abstract}
A Be/X-ray binary system known as RX J0440.9+4431 (or LS V +44 17) is a 
potential member of the uncommon gamma-ray binary class. With an orbital 
period of 150 days, this system consists of a neutron star and a Be star 
companion. The MAXI observatory discovered an X-ray outburst from the source 
in December of 2022. Early in January, the outburst reached its peak, which 
was then followed by a decrease and a subsequent rebrightening. The X-ray flux 
exceeded 1 Crab in the 15-50 keV range at this second peak. AstroSat 
observations were conducted close to the peak of the second outburst, from 
January 11 to January 12, 2023. We report here the results of our search for 
3-80 keV X-ray emission in the data of the AstroSat's LAXPC detector. It is 
found that the pulse period of the source is around 208 seconds. The source is 
found to be emitting more in the softer part of the X-ray energy range. The 
spectral characteristics can be described by employing a power-law model with 
an exponential cutoff, along with a blackbody component, interstellar absorption 
and an additional 6.4 keV iron fluorescence line.
\end{abstract}


\maketitle
\section{Introduction}
X-ray binaries (XRBs) are the peculiar X-ray systems among the most puzzling 
and brightest objects in the Universe. XRBs can be simply defined as systems 
that consist of a compact object (black hole or neutron star) and an optical 
companion (typically a normal main-sequence star) which are gravitationally 
bound to each other \cite{Manoj_2023, Nagase_1989, Liu_2007, Paul_2011}. They 
are ``close" systems due to mass transfer from the companion to the compact 
object \cite{Reig_2011}. X-ray binaries can be divided into low-mass and 
high-mass XRBs depending on the donor star's mass. In the case of low-mass 
XRBs (LMXBs) ($M \le$ 1 $M_{\odot}$), the accretion of materials happens 
through Roche-lobe overflow of the companion star \cite{Reig_2011, 
Shakura_1973}. For high-mass XRBs (HMXBs) ($M \ge$ 5 $M_{\odot}$), the 
accretion takes place through stellar wind or Be-disk of the donor star 
\cite{Manoj_2023}. They contain early-type stars as companions, such as O or B 
stars and are strong X-ray emitters. Based on the luminosity class, the HMXBs 
can be further subdivided into Be/X-ray binaries (BeXBs), where the optical 
companion can be a dwarf, giant, or sub-giant OBe star in the luminosity 
classes III-IV, and supergiant XRBs (SGXBs), in which the companion is a 
luminosity class I-II supergiant star \cite{Manoj_2023,Reig_2011}. Moreover, 
in BeXBs, the optical partners are fast-rotating stars of the type mentioned 
above with distinctive emission lines in their spectrum at some point 
of their existence (accordingly the lowercase ``e" used to denote emission) 
\cite{Reig_2011, Porter_2003, Balona_2000, Slettebak_1988}. Usually, the 
neutron star as the compact object in the binary system is considered as a 
defining characteristic of BeXBs. These non-supergiant BeXBs possess larger 
orbital periods \cite{Reig_2011, Manoj_2023}. Roughly two-thirds of HMXBs 
populations are Be/X-ray binaries \cite{Manoj_2023}. In the binary systems 
known as the accreting X-ray pulsars (XRPs), a neutron star accretes matter from the 
donor companion. The vast majority of XRPs fall under the subclass of Be/X-ray 
binaries. Here, the companion's circumstellar disk emits H$\alpha$ Balmer 
lines at least once in its life \cite{Malacaria_2022, Yan_2016, Porter_2003}. 
In BeXBs, the optical emissions predominantly come from the Be star, whereas 
the X-ray emissions provide insights into the conditions in the vicinity of 
the neutron star \cite{Yan_2016, Okazaki_2013}. BeXBs systems exhibit two 
different X-ray outburst types: Type-I and Type-II X-ray outbursts. Type-I 
commonly occurs during the periastron passage of the neutron star, while 
Type-II can occur at various orbital phases which may be linked to the
warping of the circumstellar disk's outermost region \cite{Okazaki_2013}.

RX J0440.9+4431 is a BeXB which was discovered by Motch et al.\ in 1997 
\cite{Motch_1997} during the ROSAT's survey of the galactic plane 
\cite{Salganik_2023, Manoj_2023}. Its distance was estimated to be 
$\sim$ 3.3 $\pm$ 0.5 kpc \cite{Manoj_2023}. Strong X-ray emissions ($\sim$ 10$^{34-35}$ 
erg s$^{-1}$) \cite{Salganik_2023} with pulsations (period 
$\sim$ 202.5 $\pm$ 0.5 s \cite{Reig_1999}) were discovered from this source which 
confirmed it as an X-ray pulsar \cite{Reig_1999, Victor_2023}. Reig et al.\ 
(2005) \cite{Reig_2005, Victor_2023} had done a thorough investigation of the 
characteristics of its optical counterpart. This optical companion, BSD 
24-491/LS V +44 17, was classified as a Be star which belongs to the B0.2V 
spectral class \cite{Reig_2005, Manoj_2023}. Due to the occurrence of the 
limited number of relatively dim Type-I outbursts (which is typical for BeXBs) that have been observed, the characteristics of this binary system have not 
been extensively investigated in the X-ray domain \cite{Victor_2023, 
Morii_2010, Tsygankov_2012, Ferrigno_2013}. RX J0440.9+4431 stands out as 
one of the rare X-ray binary systems where accretion continues even during 
quiescence \cite{Victor_2023, Reig_1999, LaPalombara_2012}. It is crucial to 
note that the accretion in quiescent is powered by a cold and non-ionized 
disk \cite{Victor_2023, Tsygankov_2017} along with the wind accretion.

The MAXI detector observed the first outburst from RX J0440.9+4431 during 
March 26, 2010 and April 15, 2010 \cite{Salganik_2023, Yan_2016, Morii_2010} 
with a peak luminosity of 3.9 $\times$ 10$^{36}$ erg s$^{-1}$ (3-30 
keV) \cite{Salganik_2023, Usui_2012}. \emph{Swift}/BAT also detected the next 
two minor X-ray flares following this first outburst \cite{Yan_2016, 
Krivonos_2010}. It is remarkable to note that considering the RX J0440.9+4431 
is located at a distance of 2.44$^{+0.06}_{-0.08}$ kpc \cite{Salganik_2023, 
BailerJones_2021}, the observed luminosity of the source was found to be 
remained considerably below the characteristic 10$^{37}$ erg s$^{-1}$, 
which is typically associated with Type-I outbursts in the case of BeXBs 
\cite{Salganik_2023, Morii2_2010, Tsygankov_2011, Krivonos_2010, Finger_2010, 
Reig_2011}.

The transient activity exhibited by the source was observed mainly in the 
years 2010 and 2011 \cite{Victor_2023} with luminosity peaks varying from 
(1-5) $\times$ 10$^{36}$ erg s$^{-1}$. During this span of time, the 
presence of a probable cyclotron line was reported as an absorption feature at 
$\sim$ 30 keV which denotes a magnetic field intensity of 3 $\times$ 10$^{12}$ 
G \cite{Salganik_2023, Tsygankov_2012}. However, later studies have raised 
questions about the presence of this specific spectral feature 
\cite{Victor_2023, Ferrigno_2013}. The pulse profile that was recorded 
displayed a generally simple and nearly sinusoidal pattern in 0.3-60 keV 
range. Nevertheless, some luminosity-dependent structures were evident in the 
3-15 keV range \cite{Victor_2023, Tsygankov_2012}. Usui et al.\ (2012) 
\cite{Usui_2012} noted an abrupt fall of luminosity after the initial peak. 
This was described as a consequence of the accretion stream blocking the 
emission zone. The source was most likely in a sub-critical accretion 
condition, meaning that the emission originated from a hotspot, as 
evidenced by the relatively low measured luminosity and very simple shape of 
the pulse profile \cite{Victor_2023, Basko_1976}. The soft X-ray spectra were 
also observed by various observatories. The spectra can be explained as a 
composite model of power-law and black-body components with the addition of 
the 6.4 keV iron line \cite{Salganik_2023, Usui_2012, Tsygankov_2012, 
LaPalombara_2012}.

After more than ten years of inactivity, RX J0440.9+4431 went through a 
massive Type-I outburst on December 29, 2022, which was detected by the MAXI 
all-sky monitor \cite{Nakajima_2022}. This was followed by a multi-wavelength 
campaign using NuSTAR, NICER, Chandra, \emph{Swift}-XRT, and AstroSat 
which lasted for almost four months \cite{Nakajima_2022, Mandal_2023, 
Salganik_2023, Pal_2023, Coley_2023}. \emph{Swift}/BAT data indicates that on 
January 4, 2023, the outburst peaked $\sim$ 0.6 Crab in the 15-50 keV band 
with luminosity $\sim$ 5 $\times$ 10$^{36}$ erg s$^{-1}$. The source then 
had a rapid drop in luminosity until the trend changed around January 13, 
2023, indicating the start of a significant Type-II outburst. This outburst 
achieved its peak of $\sim 2.2$ Crab on January 20, 2023, having a luminosity 
of $\sim$ 2 $\times$ 10$^{37}$ erg s$^{-1}$ in the 15-50 keV band 
\cite{Salganik_2023, Pal_2023}. The source was observed by AstroSat LAXPC on 
January 11, 2023 \cite{ISSDC}. In this paper, we report the results of the 
analysis of this observational data of RX J0440.9+4431. 

The remaining part of this paper is organized as follows. In Section 
\ref{sec2}, we provide a summary of the instrument, the observation data used 
and the data reduction procedure. The temporal and spectral analysis of the 
LAXPC data are presented in the Section \ref{sec3}.  The Section \ref{sec4} is
devoted to discuss the results of both timing and spectral analysis in more 
detail. In Section \ref{sec5}, we summarise our findings.

\section{LAXPC detectors, Observations and Data Reduction} \label{sec2}
\subsection{LAXPC}
India's first multi-wavelength astronomical observatory, the AstroSat, has 
been in active operation for about the last eight years. Launched by the Indian 
Space Research Organisation (ISRO) on September 28, 2015, it was placed in a 
circular orbit at an altitude of 650 km above the Earth's surface with a 
6$^{\circ}$ inclination and an orbital period of 98 minutes. AstroSat uses a 
collection of four equipment, which includes a UV imaging telescope 
and three X-ray sensors all of which cover a range of energies from around 
1 eV to 100 keV \cite{Roy_2021}. The hard X-ray instrument, the Large Area 
X-ray Proportional Counter (LAXPC) is one of the three X-ray sensors 
\cite{Antia_2021, Agrawal_2006, Singh_2014}. It consists of three identical 
large-area proportional counters that are co-aligned, each having multilayer 
geometry \cite{Antia_2021, Agrawal_2006, Yadav_2016a} and with a collection 
area of 100 $\times$ 36 cm$^2$ \cite{Antia_2017}. The detectors are identified 
as LX10 (or LAXPC10), LX20 (or LAXPC20) and LX30 (or LAXPC30). Each detector 
is equipped with five anode layers which are divided into seven anodes, A1-A7, 
the upper two layers each having two anodes. On each of the three sides of the 
detector, there are three veto anodes, A8-A10, as well \cite{Antia_2021}. A 
gas mixture of 90\% xenon and 10\% methane kept at a pressure of around 
2 atmospheres, fills the entire volume of LAXPC10 and LAXPC20 detectors. In 
contrast, a gas combination composed of 84.4\% xenon, 9.4\% methane and 
6.2\% argon is present in LAXPC30, which is also kept at a pressure around 
$2$ atmospheres \cite{Antia_2017}.

The LAXPC operates within the energy range of 3-80 keV. The lower energy 
threshold of this set of devices is influenced by the negligible transmission 
of X-ray (below 3 keV) through 50 $\mu$m thick aluminized mylar. On the 
other hand, the reduced detection efficiency at higher energy leads to the 
higher energy threshold \cite{Agarwal_2017}. The LAXPC possesses the 
capability to study fluctuations in kHz scale at a high counting rate. A large 
photon collection area with robust detection efficiency over the complete 
energy range is necessary for this. The effective area of all three units is 
$\sim$ 4500 cm$^2$ at $\sim$ 5 keV, $\sim$ 6000 cm$^2$ at $\sim$ 10 keV and 
$\sim$ 5000 cm$^2$ at $\sim$ 40 keV \cite{Agarwal_2017, Roy_2019}.

This instrument's main objectives are broadband spectrum observations and 
high-time-resolution research to examine rapid changes in intensity. This is 
accomplished by the instrument's ability to precisely identify each observed 
X-ray photon's arrival time, with an accuracy of 10 $\mu$s. Additionally, it 
uses a $1024$-channel pulse-height analyser to calculate each photon's energy 
\cite{Roy_2019}. Agrawal (2006) \cite{Agrawal_2006} and Singh et al.\ (2014) 
\cite{Singh_2014} both provide detailed descriptions of the AstroSat 
observatory and its equipment. Yadav et al.\ (2016) \cite{Yadav_2016a}, 
Agrawal et al.\ (2017) \cite{Agarwal_2017}, and Roy et al.\ (2016) 
\cite{Roy_2016} give a more thorough explanation of the characteristics of the 
LAXPC instrument. Antia et al.\ (2017) \cite{Antia_2017} provides comprehensive 
information on calibration specifications.

\subsection{AstroSat Observation}
\begin{figure}[h!]
\includegraphics[scale=0.8]{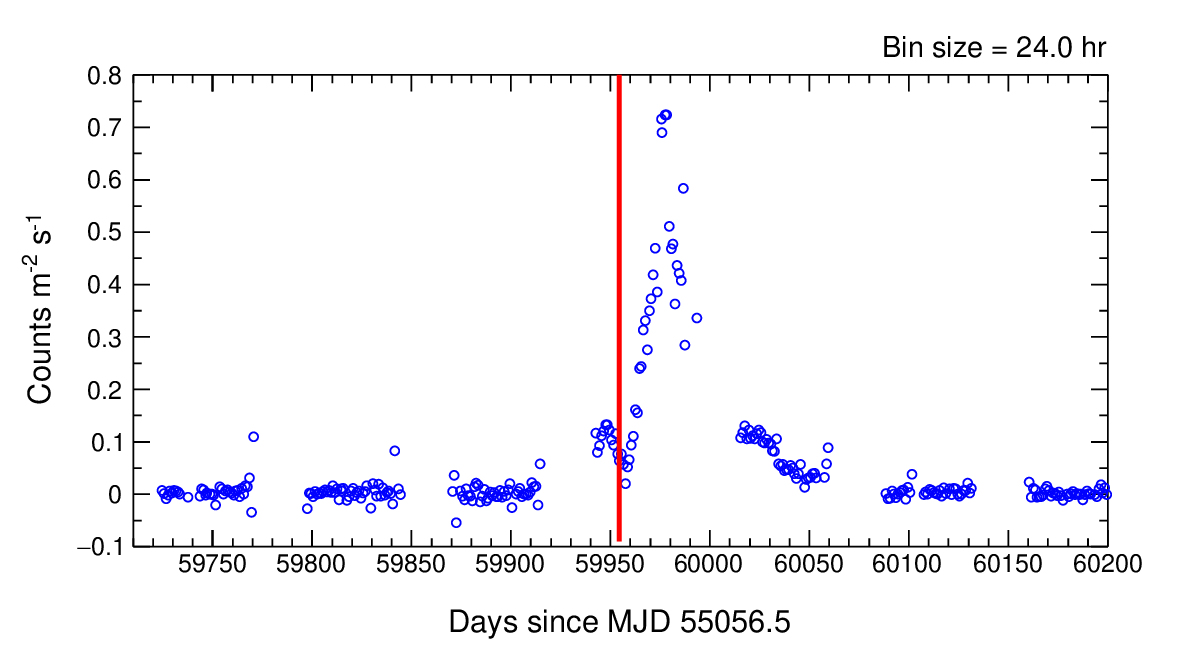}
\vspace{-0.2cm}
\caption{The light curve of the RX J0440.9+4431 observed by MAXI/GSC during the 
period of 2022 to 2023 \cite{Morii_2010, web_maxi}. A significant and 
substantial outburst event was identified emanating from RX J0440.9+4431 
within the energy range of 2 to 20 keV. The vertical strip indicates 
the specific time period of the AstroSat observations.}
\label{fig1}
\end{figure}

AstroSat monitored the source RX J0440.9+4431 from 11:52:07.64 UT on January 
11, 2023 to 00:43:20.53 UT on January 12, 2023, covering its eight orbits. 
In terms of Modified Julian Day (MJD) this observation of AstroSat lies within 
the MJD period 59955.49 to 59956.03. Background data used for this source 
was collected between May 2, 11:55:52.52 UT to May 3, 09:50:52.35 UT in the 
year 2022. This background data was obtained by pointing the instrument at a 
region devoid of any sources. The background data have been used for obtaining 
the source's background-subtracted light curve. We obtain all these data from 
the archival of the AstroSat in the Indian Space Science Data Centre (ISSDC) 
\cite{ISSDC}. It should be mentioned that between 2022 and 2023, MAXI/GSC 
detected a strong X-ray burst from RX J0440.9+4431 in the 2 to 20 keV energy 
range \cite{Morii_2010}. Fig.\ \ref{fig1} shows the source's one-day binned 
MAXI/GSC light curve in 2-20 keV from MJD 59500.0 to MJD 60200.0 
\cite{Morii_2010, web_maxi}. It is coupled with the AstroSat observation on 
MJD 59955.49 as mentioned above, denoted by a thick red vertical line. The 
source was about to have a Type-II outburst when the AstroSat observation was 
made, as seen from the MAXI/GSC light curve.

\subsection{Data Reduction}
The data reduction process wass carried out using the LAXPC data reduction 
pipeline software ``LAXPCsoftware22Aug15". The ``lxplevel2datapipeline" utility 
was used to convert Level-1 raw data files into Level-2 data. Data was 
analysed for each of the three LAXPC units individually. The Level-2 data 
consists of light curve, information about each detected X-ray photon, pulse 
height and the originating anode layer in (i) broadband counting mode (modeBB),
(ii) event mode (modeEA), and (iii) mkf files (which stores housekeeping data 
and parameter files). Using the LaxpcSoft standard tasks, the light curves and 
energy spectrum were obtained from the Level-2 event file. The software 
routine generates the Good Time Interval, which contains timing details for 
Earth occultation and the South Atlantic Anomaly. We utilise average 
background counts to determine the source light curve with the background 
subtraction process. In addition to background correction, to account for the 
variations in the arrival times of the photons due to the motion of the Solar 
system, these photon arrival times have undergone barycentric correction to 
the Solar system barycenter using the AstroSat barycentric correction utility 
``as1bary". It requires HEASOFT software \cite{HEASOFT} package version 6.17 or 
higher. We use the latest version (6.30.1) available at the time of our 
analysis. HEASOFT incorporates various tools and programs. XRONOS tool 
\cite{Stella_1992} was used for the timing analysis. FTOOLS \cite{FTOOLS} were 
primarily used for general data extraction and analysis. The XSPEC package 
\cite{Arnaud_1996} played a key role in spectral analysis. We use data 
only from the LAXPC20 detector as LAXPC30 was impacted by gas leakage and 
hence turned off on March 8, 2018 due to anomalous gain fluctuations. We did 
not include the data from LAXPC10 as it was operating at low gain 
\cite{Jitesh_2019, Chandra_2020}. We pick the 3-30 keV energy band for the 
spectrum since the background is dominant above the 30 keV. 

\section{Analysis of Data}\label{sec3}
\subsection{Timing Analysis}
\begin{figure}[h!]
\includegraphics[scale=0.48, angle=270]{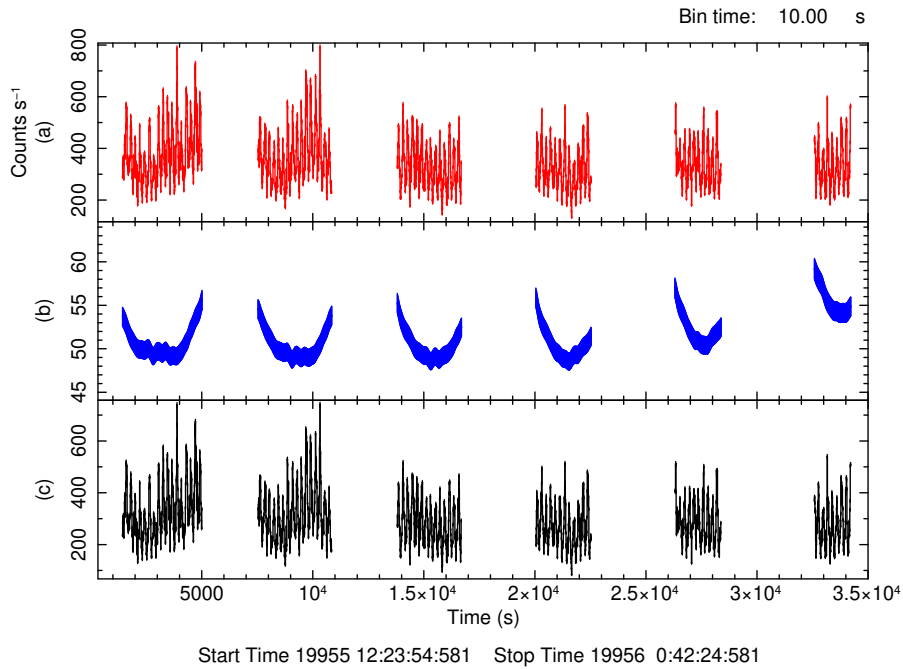}
\caption{(a) Light curve of the X-ray source RX J0440.9+4431 
including the background count rates. (b) Background light curve. 
(c) Background subtracted light curve of the source. The gaps in the light 
curve is due to the passage of the satellite through the South Atlantic 
Anomaly regions. In these plots, 10 s binning is used in the energy range of 
3-80 keV.}
\label{fig2}
\end{figure}

The light curve of the source was produced from the above-mentioned AstroSat
data, i.e.\ the data obtained from all the observations of LAXPC20 by using 
the 10 seconds averaged count rates in the 3-80 keV band, which is presented 
in the top panel of Fig.\ \ref{fig2}. To have the best signal-to-noise ratio, 
we extract data only from the top layer of the LAXPPC20 instrument. As 
mentioned earlier, there are eight orbits of observation and for those 
combined orbits, we also generate the corresponding backgrounds shown in the 
middle panel of Fig.\ \ref{fig2}. The background varies between $\sim$ 48 - 61 
counts s$^{-1}$ during the observations. To get the final light curve, 
this background was subtracted from the light curve using the ``lcmath" tool. 
The background-subtracted light curve is shown in the bottom panel of Fig.\ 
\ref{fig2}. For clarity in the visibility of pulsations the first segment 
of the light curve of Fig.\ \ref{fig2} is also shown in Fig.\ \ref{fig3}. 
Pulsations of X-rays from the source are distinctly visible in the light curve
from both figures.  

Employing the FTOOL subroutine ``powspec" in HEASOFT, the one-second binned 
power density spectrum (PDS) of the source was generated using the light curve 
of the source given by the LAXPC20 detector. We divide the light curve into 
stretches of 1024 bins per interval. The final PDS was created by averaging 
the PDSs from all the segments. The generated PDS has a very significant peak 
at $\sim$ 0.005 Hz corresponding to the source's spin period of $\sim$ 200 s 
(see Fig.\ \ref{fig4}). The PDS shows no quasi-periodic oscillation (QPO) 
\cite{Reig_2011} peaks since no additional nearby significant peaks have been 
observed except few harmonics in the spectrum. Applying the standard $\chi^2$ 
maximization approach with the FTOOLS job ``efsearch" a more accurate 
estimation of the pulse period can be obtained. The period was searched around 
200 s as approximated in the PDS generated from the light curve. It was found 
to be 207.93 s (see Fig.\ \ref{fig5}) confirming the pulse period detected by 
the NICER mission \cite{Manoj_2023}. 

\begin{figure}[h!]
\includegraphics[scale=0.42, angle=270]{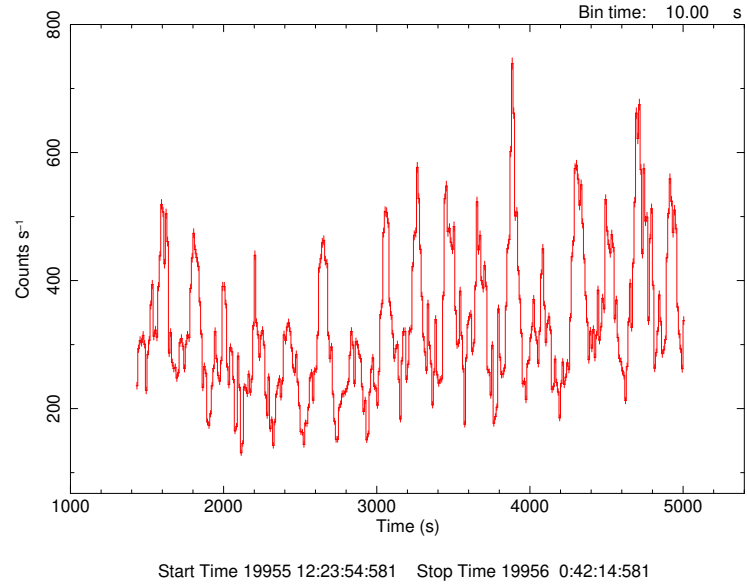}
\caption{\label{fig3} Time scale amplified background-subtracted light curve 
of the source RX J0440.9+4431 in 3-80 keV energy range obtained by using 
the 10 s bins.}
\label{fig3}
\end{figure}

\begin{figure}[h!]
\includegraphics[scale=0.42, angle=270]{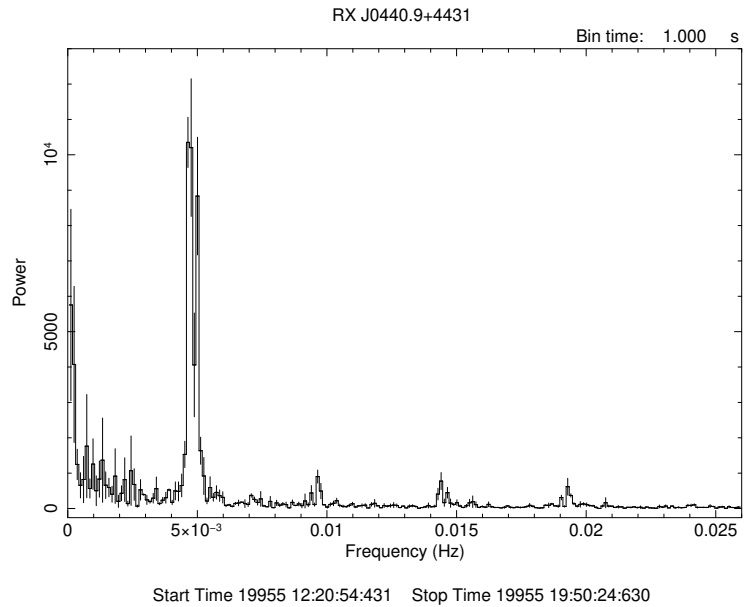}
\caption{The power density spectrum of RX J0440.9+4431 generated by using the
LAXPC20 data from 10 s binned barycenter corrected light curve in 3-80 keV 
energy band. A strong pulsation with a peak at $\sim$ 0.005 Hz was found in 
the PDS.}
\label{fig4}
\end{figure}

\begin{figure}[h!]
\includegraphics[scale=0.42, angle=270]{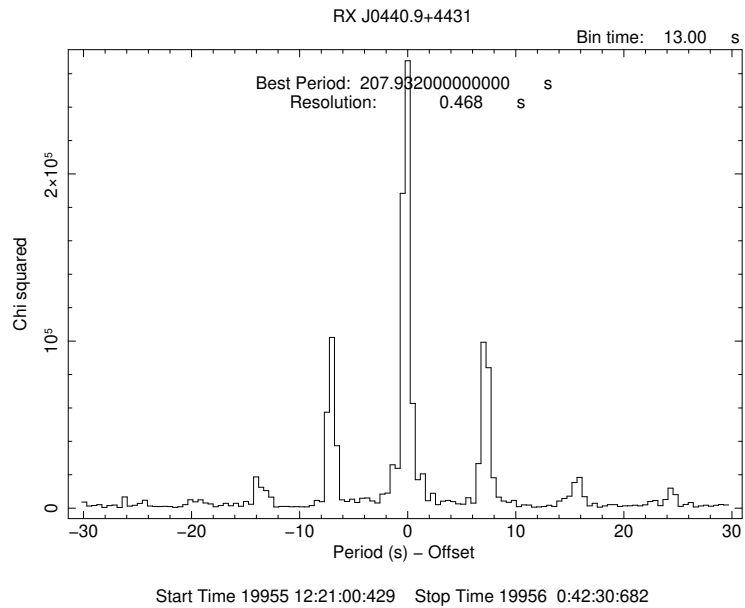}
\caption{The pulse period of the pulsar RX J0440.9+4431 estimated by using the
FTOOLS subroutine ``efsearch" from the LAXPC20 data's light curve. The period 
was found to be around 207.932 s.}
\label{fig5}
\end{figure}

\begin{figure}[h]
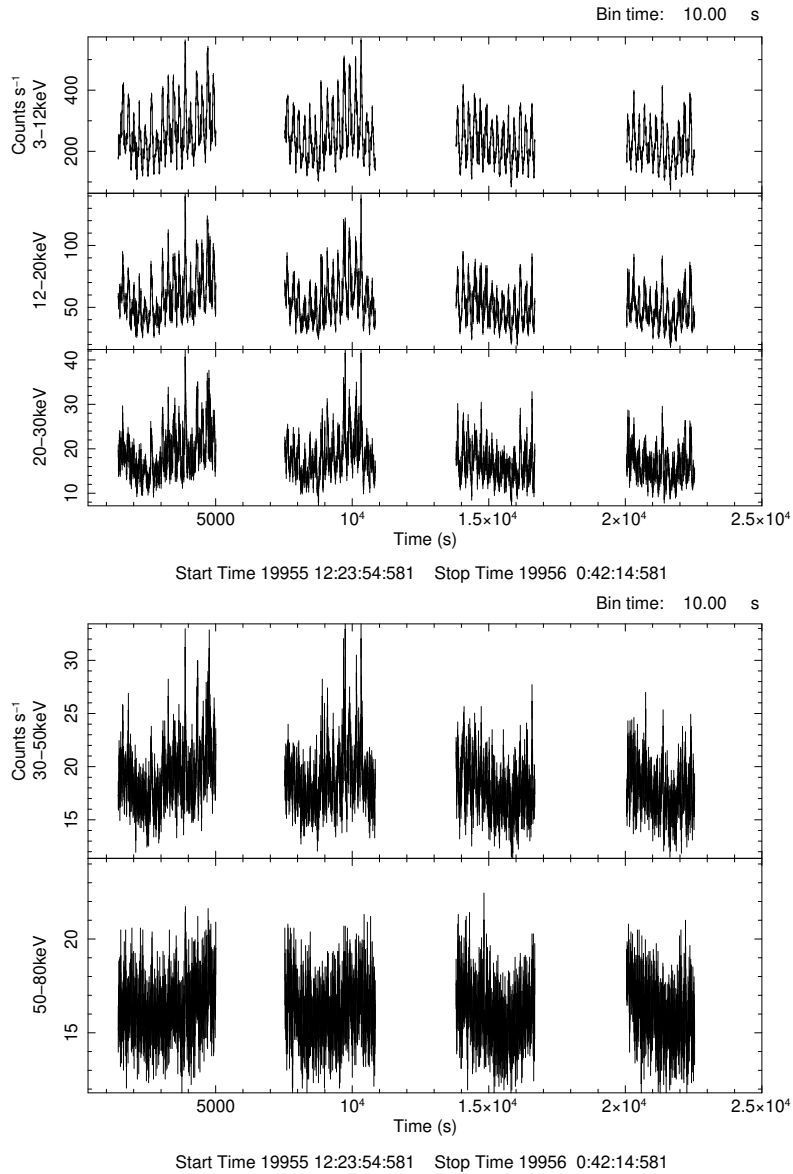

\includegraphics[scale=0.42, angle=270]{2_lc_L1_3-12_12-20_20-30keV_zoomed.ps}
\includegraphics[scale=0.42, angle=270]{2_lc_L1_30-50_50-80ke_zoomed.ps}
\caption{Background subtracted light curves for (a) 3–12 keV, (b) 12–20 
keV, (c) 20-30 keV, (d) 30–50 keV and (e) 50–80 keV for LAXPC20 data of
the pulsar RX J0440.9+4431.}
\label{fig6}
\end{figure}

To investigate how intensity fluctuations depend on energy, light curves were 
produced within five unique energy bands: 3-12 keV, 12-20 keV, 20-30 
keV, 30-50 keV and 50-80 keV for the LAXPC20 detector. Only Layer 1 data 
were utilised for the construction of the light curves in the energy range from 
3 to 20 keV. This selection was considered due to the fact that most 
incident photons ($\sim$ 90 \%) with energy $<$ 20 keV are likely to be 
absorbed in Layer 1 \cite{Roy_2019}. Similar to this, light curves in the 
20 to 50 keV range were produced by summing data from Layers 1 and 2, 
while those in the 50 to 80 keV range used data from all five layers. The 
count rates for the LAXPC20 detector within these five energy bands are shown 
in Table \ref{tab2}. Fig.\ \ref{fig6} shows that the higher amplitudes of 
oscillations are noticeable in the energy bands of 3-12 keV, 12-20 keV and 
20-30 keV than in the 30-50 keV and 50-80 keV bands.

\begin{figure}[h]
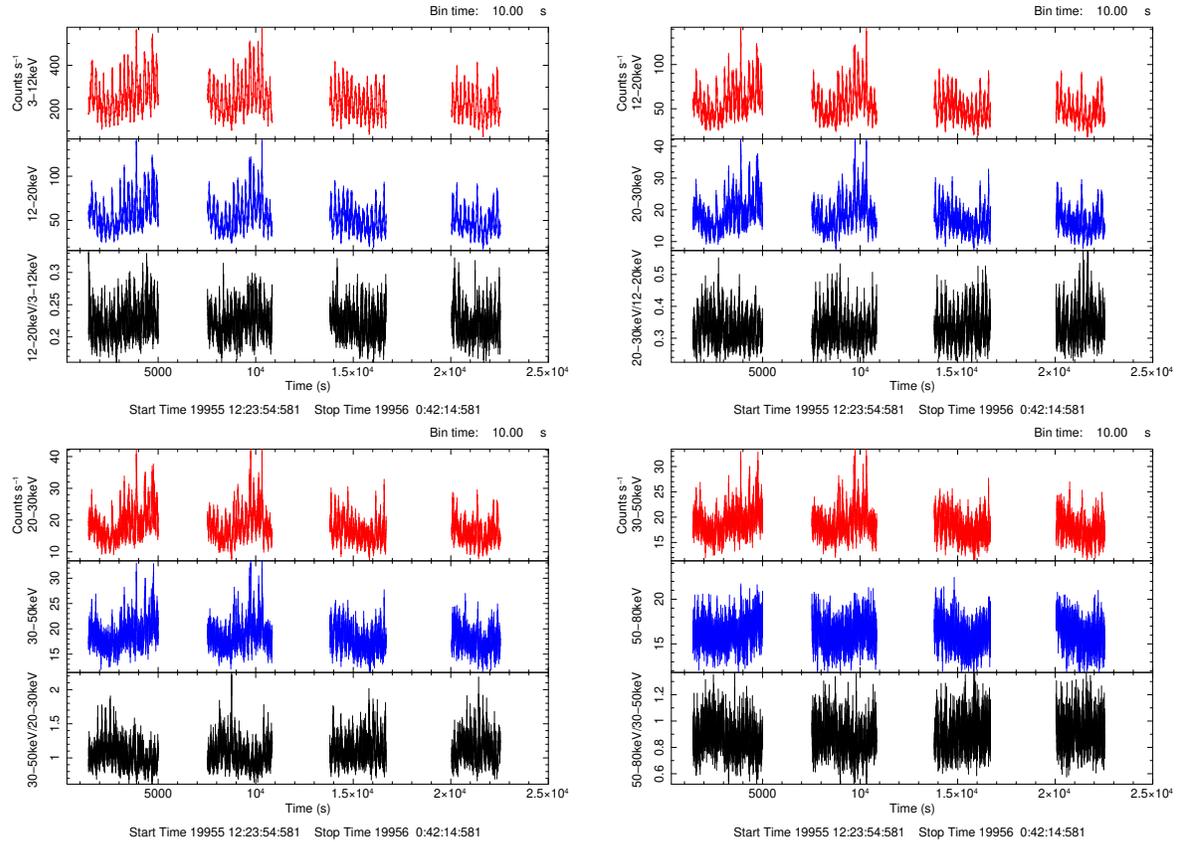

\includegraphics[scale=0.3, angle=270]{2_hr_12-20_7-12_10s.ps}
\includegraphics[scale=0.3, angle=270]{2_hr_20-30_12-20_10s.ps}
\includegraphics[scale=0.3, angle=270]{2_hr_30-50_20-30_10s.ps}
\includegraphics[scale=0.3, angle=270]{2_hr_50-80_30-50_10s.ps}
\caption{Hardness ratios for 12-20 keV/7-12 keV (top left panel), 
20-30 keV/12-20 keV (top right panel), 30-50 keV/20-30 keV (bottom left panel) 
and 50-80 keV/30-50 keV (bottom right panel). The bottom section of each panel 
(black color) represents the hardness ratio of the corresponding energy band.}
\label{fig7}
\end{figure}

To analyse the spectral evolution during the intensity oscillations, hardness 
ratios (HRs) were computed. These HRs are defined as the count rates in a 
higher energy band divided by the count rates in a lower energy band. We 
generate four hardness ratio diagrams (background subtracted) for 12-20 
keV/3-12 keV, 20-30 keV/12-20 keV, 30-50 keV/20-30 keV and 50-80 keV/30-50 keV.
In Fig.\ \ref{fig7}, it is noteworthy to notice that the source's intensity 
shows no correlation with the HR. It indicates that the intensity oscillations 
are not directly related to spectral changes within the detected energy ranges.

\begin{center}
\begin{table}[h!]
\caption{Background subtracted LAXPC20 count rates in five energy bands of the 
source RX J0440.9+4431.}
\vspace{8pt}
\begin{tabular}{c c}
\hline
\hline
 Energy Range & Average Count Rate \\
       (keV)        & (counts s$^{-1}$) \\
\hline
 3-12         & 240.41 \\
 12-20        & 53.57  \\
 20-30        & 17.75  \\
 30-50        & 18.88  \\
 50-80        & 16.73  \\
\hline
\end{tabular}
\label{tab2}
\end{table}
\end{center}

\subsection{Spectral Analysis}
Since the spectral response matrix of the LAXPC20 detector is well determined, 
we only conduct a thorough spectral analysis on LAXPC20 data. The spectral 
fitting and statistical analysis are done using the XSPEC version 12.12.1. 
Data from all the eight orbits observed by LAXPC20 were merged to derive the 
spectrum of the source in the energy band ranging from 3 keV to 80 keV. To 
maximize the inclusion of X-ray photons and to minimize the background, the 
spectrum was generated only from the top layer of LAXPC20. The background 
spectrum was found to be dominating above 30 keV. Therefore, the photons below 
3 keV (ignored channel: 1) and above 30 keV (ignored channels: 30-77) for 
LAXPC are ignored to avoid larger systematic errors. The LAXPC software has a 
single routine to generate the source spectra, the background spectra and the 
response matrix files. The spectra were fitted with a combination of models 
``TBabs*(cutoffpl + bbodyrad + gaussian)” which are built-in models found in 
XSPEC. 

\begin{figure}[h!]
\includegraphics[scale=0.42, angle=270]{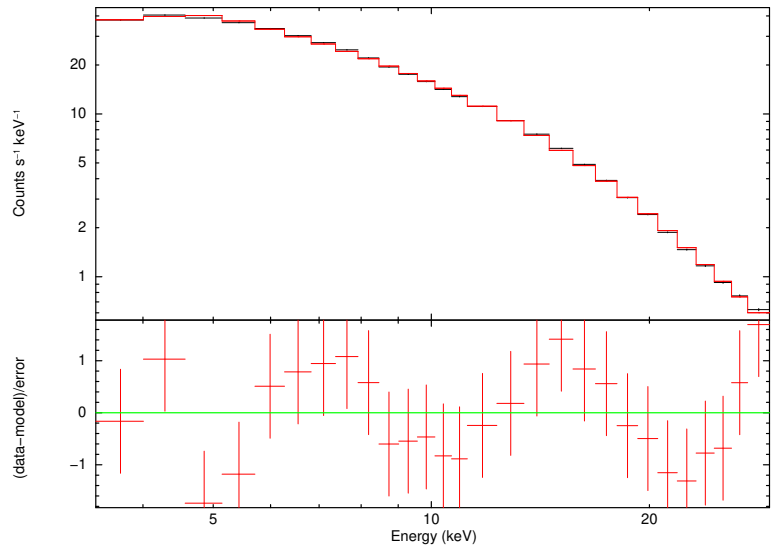}
\caption{Fitted LAXPC spectrum of the source RX J0440.9+4431. The best fit
is obtained by using the model TBabs*(cutoffpl + bbodyrad + Gaussian). The 
fitted model spectrum is shown by the red solid line along with the observed
one in black colour. The residuals between the data and the corresponding 
model predictions are shown in the lower section of the plot.}
\label{fig8}
\end{figure}

To model the spectrum, we first consider a single-component model, the 
power-law model with an exponential cutoff (cutoffpl). Even after using a 
systematic error of 0.02 (i.e. 2\%) in data, it gives a poor fit with
$\chi^2$/dof value of 5056/25. 
Next, we add a blackbody radiation model (bbodyrad) to the power-law model. 
With this addition, we gain a noticeable improvement in fitting parameters 
($\chi^2$/dof = 2723.11/23). However, the combined model left a residual at 
$\sim$ 6.4 keV. Therefore a Gaussian line at 6.4 keV (due to iron K$\alpha$ 
line) was included which significantly improved the fit with $\chi^2$/dof = 
270.98/20. Finally, we include the multiplicative model, TBabs 
\cite{Wilms_2000, Misra_2020} for interstellar absorption. Here the column 
density value has been fixed at $N_H =$ 1.0 $\times$ 10$^{22}$ cm$^{-2}$. It 
gives a very good fit of $\chi^2$/dof = 22.29/20 (i.e. $\chi^2_{red}=1.11$). 
Table \ref{tab3} lists the spectral parameters obtained from the best fit of
our combined model. The best-fitted spectrum of the source is shown in 
Fig.\ \ref{fig8}.

\begin{center}
\begin{table}[h]
\caption{Best fitting spectral parameters obtained from the model: 
TBabs*(cutoffpl + bbodyrad + Gaussian).}
\vspace{8pt} 
\begin{tabular}{c p{1.4cm} p{2.8cm} c c}
\hline \hline
\vspace{4pt}
& Model  & Parameter & Value \\
\hline
& TBabs      & $N_H$ ($10^{22}$ cm$^-2$)       & 1.0000  &\\
 
& cutoffpl   &  PhoIndex                       & 0.9495  &\\
 
&            &  HighECut (keV)                 & 19.5063 &\\ 

&            &  Norm                           & 0.1032  &\\

& bbodyrad   &  kT (keV)                       & 0.2818  &\\

&            &  Norm ($\times 10^{5}$)         & 1.9452  &\\

& Gaussian   &  LineE ($\times 10^{-15}$ keV)  & 9.2099  &\\
 
&            &  Sigma (KeV)                    & 4.4204  &\\

&            &  Norm                           & 0.1855  &\\ 

\hline
\end{tabular}
\label{tab3}
\end{table}
\end{center}                  
	
\section{Results and Discussion} \label{sec4}
The AstroSat observation of RX J0440.9+4431 provides us with an opportunity to 
investigate the behaviour of the pulsar at X-ray energies 3-80 keV during 
the enormous outburst in 2023. \emph{Swift}/BAT observations show that the 
huge outburst reached a record-high flux of 2.3 Crab \cite{Manoj_2023}. During 
the outburst of the source, cleaned event data were used to create a 10-second 
binned light curves in various energy ranges to observe the periodicity in the 
data. According to Reig \& Roche in 1999 \cite{Reig_1999}, the source RX 
J0440.9+4431 has been classified as a persistent BeXB due to its longer spin 
period and less X-ray variability than typical transient pulsars. In such 
cases, the neutron star and the Be star are in a wide orbit. The neutron star 
orbits through the outer portion of the Be circumstellar disc, where matter 
density is low. However, RX J0440.9+4431 displayed an eruption which was 
contrary to what was anticipated. This suggests that the source was in 
transition to a supercritical regime \cite{Victor_2023, Usui_2012}. 

The light curves of the source clearly exhibit the prominent intensity 
oscillations of a period of approximately 200 seconds. These oscillations are 
readily visible across all of the light curve data. The $\sim$ 200 s X-ray 
pulsations are clearly visible also in the energy-resolved light curves in 
3-12 keV, 12-20 keV, 20-30 keV, 30-50 keV and 50-80 keV bands. To search for 
the periodic signals, the barycentre corrected event data were employed and 
efsearch task in FTOOLS was executed. This task folded the light curve over a 
trial period and identified the most likely period. From this analysis, we 
obtain a measurement of the pulse period $\approx$ 207.9 s which confirms the 
pulse period as described by NICER observation for this source 
\cite{Manoj_2023}. The evolution of the pulse period could not be studied due 
to AstroSat's observation for a very short period of time. However, Reig \& 
Roche in 1999 \cite{Reig_1999} found its period to be $\sim 202.5$ s. 
According to LaPalombara (2012) \cite{LaPalombara_2012}, the pulse period 
was $\sim 204.96$ s. Another study showed its period to be $\sim 205$ s 
\cite{Usui_2012}. Thus, our measured period of 207.9 s in the recent AstroSat 
observation is notably longer than the one recorded in 1999. This discrepancy 
suggests that over the previous ten years, the neutron star has been slowing 
down. This deceleration indicates that the neutron star's rotation rate will 
gradually slow down over time. 

In energy-resolved light curves, the spectral changes can be seen clearly. It 
is noticed that the source is softer, i.e., it emits more low energy photons, 
primarily in the 3-12 keV band ($\sim$ 240 counts s$^{-1}$ on average). This 
may refer to cold disk accretion in our Be/X-ray binary. Be/X-ray binaries are 
thought to be persistent systems. This hypothesis supports that the long-term, 
low-level X-ray activity of these sources is probably due to the accretion of 
matter onto the neutron star from the winds of the star. BeXBs sometimes show 
a periodic increase in X-ray activity rather than being persistent X-ray 
generators. This is thought to be caused by changes in the size of the 
equatorial disc of the Be star \cite{Reig_2011}. During the outburst of such 
systems, the compact object, i.e., the neutron star and the companion star's 
equatorial disk comes into proximity (periastron passage). Thus we get 
quiescent emission from the source \cite{Rothschild_2012}.

During the modelling of the RX J0440.9+4431 spectrum, it does not show any 
presence of narrow absorption features which can be identified as a cyclotron 
line as reported in earlier studies \cite{Salganik_2023, Tsygankov_2012}. At 
this luminosity of 2 $\times$ 10$^{37}$ erg s$^{-1}$ (as observed by 
\emph{Swift}/BAT) during Type-II outburst in 2023, opting for a power-law 
model with an exponential cutoff was found to be a feasible choice 
\cite{Salganik_2023, Coburn_2002}. It is a characteristic of high accretion 
rates \cite{Salganik_2023, Becker_2007, Farinelli_2016}. The Fe K$\alpha$ line 
found at 6.4 keV may be generated through the interaction of the hard X-ray 
radiation with relatively cooler matter. This line may be associated with 
iron atoms in a nearly neutral state (composite of Fe II and Fe XVIII) 
\cite{Manoj_2023, Reig_2013, Ebisawa_1996, Liedahl_2005}.  

\section{Conclusion} \label{sec5}
Here, using the AstroSat LAXPC data, we present the findings of the brief 
investigation of the temporal and spectral properties of the BeXB system 
RX J0440.9+4431 during a giant outburst observed at the beginning of 2023. A 
shift from the subcritical to supercritical accretion regime can be inferred 
from the change in the intensity and flux hardness during the outburst. The 
pulse period of the source was also estimated and it was found to be prominent 
at 207.9 seconds. A longer pulse time has been observed as compared to the 
previous measurements, which suggests that the rotation of the neutron star 
has slowed down during the past ten years. A study of broadband spectra near 
the outburst was also carried out. The spectra of the source can be modelled 
using interstellar absorption along with high energy cutoff power-law and 
blackbody radiation which shows the presence of no narrow absorption features, 
such as cyclotron lines. The iron line at 6.4 keV shows the presence of nearly 
neutral iron atoms in the interaction of hard X-ray with colder matter.

Overall, our investigation of the outburst of RX J0440.9+4431 reveals important 
information on the complicated dynamics of the Be/X-ray binary during severe 
events and delivers insightful knowledge about its behaviour and spectral 
properties.

\section{Acknowledgemnet} \label{sec6}
We express our gratitude to the LAXPC team for their contribution to the 
development of the LAXPC instrument. In this study, we used the data obtained 
through the HEASARC Online Service, which was made available by NASA/GSFC, in 
support of NASA High Energy Astrophysics Programs. This paper also makes use 
of data from the AstroSat mission of the Indian Space Research Organisation 
(ISRO), which is archived at the Indian Space Science Data Centre (ISSDC). UDG 
is thankful to the Inter-University Centre for Astronomy and Astrophysics 
(IUCAA), Pune, India for awarding the Visiting Associateship of the institute.

\bibliographystyle{apsrev}

\end{document}